\documentclass[12pt,a4paper]{article}
\usepackage{graphicx}
\usepackage{times}
\usepackage{amssymb,amsbsy}
\usepackage{amsmath}
\newcommand{\msun}{$M_\odot$\,}

\newcommand{\as}{\prime\prime}
\def\fr{\frac}

\newcommand{\barr}{\begin{array}}
\newcommand{\earr}{\end{array}}
\newcommand{\berr}{\begin{eqnarray}}
\newcommand{\err}{\end{eqnarray}}
\newcommand{\berrno}{\begin{eqnarray*}}
\newcommand{\errno}{\end{eqnarray*}}
\newcommand{\be}{\begin{equation}}

\newcommand{\pol}[1]{\stackrel{\rm LCP}{\mathrm{RCP}}}

\newcommand{\no}{\nonumber}

\renewcommand{\a}{\alpha}

\renewcommand{\l}{\lambda}

\newcommand{\s}{\sigma}

\title{\bf Gravitational Lensing and Microlensing in Clusters: Clusters as Dark Matter Telescopes}
\author{Margarita Safonova\thanks{E-mail:  margarita.safonova62@gmail.com}\\
\normalsize $^1$ Indian Institute of Astrophysics, Bangalore 560034, India}

\date{\mbox{}}
\begin{document}
\maketitle
\pagestyle{empty}
%
%
\def\bull{\vrule height .9ex width .8ex depth -.1ex}
\makeatletter
\def\ps@plain{\let\@mkboth\gobbletwo
\def\@oddhead{\it Bulletin de la Soci\'{e}t\'{e} Royale des Sciences de Li\`{e}ge, Vol. 87, Actes de colloques, 2018, pp. 360-364}\def\@oddfoot{\hfil\scriptsize\bull\quad
First Belgo-Indian Network for Astronomy \& Astrophysics (BINA) workshop, held in Nainital (India), 15-18 November 2016 \quad\bull}%
\def\@evenhead{\it Bulletin de la Soci\'{e}t\'{e} Royale des Sciences de Li\`{e}ge, Vol. 87, Actes de colloques, 2018, pp. 360-364}\let\@evenfoot\@oddfoot}
\makeatother
%
%
\def\beginrefer{\section*{References}%
\begin{quotation}\mbox{}\par}
\def\refer#1\par{{\setlength{\parindent}{-\leftmargin}\indent#1\par}}
\def\endrefer{\end{quotation}}
%
%
{\noindent\small{\bf Abstract:} 

Gravitational lensing is brightening of background objects due to deflection of light by foreground sources. Rich clusters of galaxies are very effective lenses because they are centrally concentrated. Such natural Gravitational Telescopes 
provide us with strongly magnified galaxies at high redshifts otherwise too faint to be detected or analyzed. With a lensing boost, we can study galaxies shining at the end of the ``Dark Ages". We propose to exploit the opportunity provided by the large field of view and depth, to search for sources magnified by foreground clusters in the vicinity of the cluster critical curves, where enhancements can be of several tens in brightness. Another aspect is microlensing (ML), where we would like to continue our survey of a number of Galactic globular clusters over time-scales of weeks to years to search for ML events from planets to hypothesized central intermediate-mass black holes (IMBH).
}

%
%
\section{Introduction}

Clusters of galaxies and clusters of stars have one thing in common: a substantial part of their mass is dark. 
In clusters of galaxies, it is the non-barionic dark matter that constitutes upto 90\% of all cluster's mass. In 
clusters of stars, and here we consider only globular clusters (GC), 
it is the barionic dark matter in the form of black holes, neutron stars, brown dwarfs, and planets. The dark 
matter is by definition dark, it does not shine and interacts only gravitationally, thus the only efficient way 
for its detection is through its gravitational effect on light --- by gravitational lensing (GL) --- a phenomenon where the compact mass acting as a lens distorts the surrounding space-time which results  
in the deflection of light from background objects creating false images. GL probes the mass directly, therefore dark 
matter is not hidden to the GL. Additional bonus is the amplification of the light from background sources affected 
by the gravitational lens. The amplification, whether in strong-lensing regime or as a transient -- microlensing regime,
`boosts' the background sources that may otherwise be too dim or far away to be visible. Dark matter in galaxy clusters 
magnifies faint background sources making them visible to us (as high-redshift galaxies). Dark matter in globular clusters 
lenses background stars making itself visible to us (as planets, black holes, neutron stars). 

We propose to search for high-redshift sources magnified by the  foreground lensing clusters, where combination of adaptive optics and gravitational magnification offers great opportunities. We also intend to continue our program of searching for ML events from different range of lens masses in Galactic GCs. Such a study would be useful to obtain information relating to mass function of globular clusters.

\section{Strong lensing by clusters of galaxies}

Early population of star-forming galaxies is intrinsically very faint, low-mass, and of low extent. One effective way to 
find early galaxies is by using distortion and magnification by Dark Matter Telescopes! Light of distant sources (galaxies) 
passing through the foreground galaxy cluster bends and is amplified. Rich clusters of galaxies at redshifts $z \sim $2 with masses of order $10^{14}\,$ \msun are very effective lenses because they are centrally concentrated. Such natural gravitational telescopes provide us with strongly magnified galaxies at (very) high redshifts which otherwise are too faint to be detected or analysed in the unlensed state. The magnification in some clusters can reach a factor of 25-50 (Ellis et al. 2001). Strong lensing fields are a factor 
$\sim$ 5-10 more efficient than blank fields of the same size for high-redshift ($>7$) galaxy searches (e.g. Pell\'{o} et al. 2005).  

\subsection{Lenses and search technique}

The methodology starts with identification of a sample of clusters at intermediate redshifts, whose mass distributions are well-constrained by arcs and multiple images of known redshifts. Next is the search along the critical curves of the cluster. Sources that lie on or near the caustics (one-dimensional loci of formally infinite magnification) 
in the source plane are imaged on or in the vicinity of the critical curves (lens plane images of the caustics) with very high factor. Location of critical lines depends on cluster's mass distribution, angular diameter distances between lens, source and observer, and $z$ of the source, thus is calculable once the mass distribution of the cluster lens is constrained. For sources occupying a specific redshift range, e.g. $2<z<7$, particularly high magnification can be expected ($\sim$ 40-fold), and they are detectable in optical bands. At $z>7$, the most relevant signatures are expected in the near-IR $\l \ge 1\,\mu$m (due to shift of Lyman-$\a$ emission). By only searching say $z > 5$ critical line, we minimize contamination from lower redshift sources.

\subsection{Large telescopes advantages}

Large field of view combined with a good angular resolution ($0.^{\as}4$) provides an advantage in efficient imaging the regions in and around the image plane critical curves
(which are typically $\sim$ tens of arcseconds in extent). Multi-waveband observational capabilities will provide an aid to identifying images on the basis of colour, facilitating separation of foreground objects (including the lens cluster members) from the background objects. Generating deep exposure by stacking successive observations will enable us to go down to faint image magnitudes of around 25 -- 26, accessing source magnitudes that are several times fainter. Spectroscopic follow-up will allow accurate determination of the redshifts of the images, and aid in identifying and separating possible blends.

\section{Microlensing in globular clusters}

ML is characterized by the transient, achromatic brightening of a background star due to gravitational deflection of its light by a massive lens passing near or intersecting our line of sight to the source. This results in the distortion of the source into multiple images whose total brightness is greater than that of the original source. The temporal dependence of the event results in an observable light curve, from which inferences regarding the ML event are made, including a probabilistic estimate of the mass of the lens. GCs are ideal targets to search for the ML signatures as the probability of lensing is high in the dense cluster environment, distances, kinematics of lenses and sources is generally well constrained. GCs are easily accessible by ground-based telescopes, and compactness allows photometry of thousands of stars in one CCD frame.

\subsection{Intermediate-mass black holes (IMBH)}

Globular clusters are hypothesized to harbour intermediate-mass ($10^3-10^4$ \msun) black holes in their centres. Gravitational ML of a background star by the central IMBH is the only currently existing tool that can resolve the nature of the central dark mass, as there is a significant difference in the lensing signatures of a single point-mass lens (a single black hole) and an ensemble of point-mass lenses (which would be if the central mass consisted of many low-mass objects). In 2010, our group initiated monitoring of a set of selected GCs looking for ML signature of possible central IMBH, which would act as a lens, amplifying 
light of stars passing behind (Safonova \& Stalin, 2010). Since the time scale of such event is of the order of $\sim $yrs, the monitoring cadence is once a month. We use differential photometry to search for variability in our data using the package originally developed by Wozniak (2000) and modified by W.~Pych (2012). Differential photometry is unaffected by blending and, more importantly, is sensitive to ML events due to stars that are too faint to be detected at baseline, which is usual for the cores of globular clusters. The reference frame (RF) is matched to the seeing of each science image and subtracted to obtain a series of difference images. The GC centre is examined for variability and  candidates transient/variable are further studied after constructing the light curves. In Fig.~1 we present the sample data from the globular cluster M13, demonstrating the technique (Joseph et al. 2015).  

\begin{figure}[h]
\begin{minipage}{7cm}
\centering
\includegraphics[width=8cm]{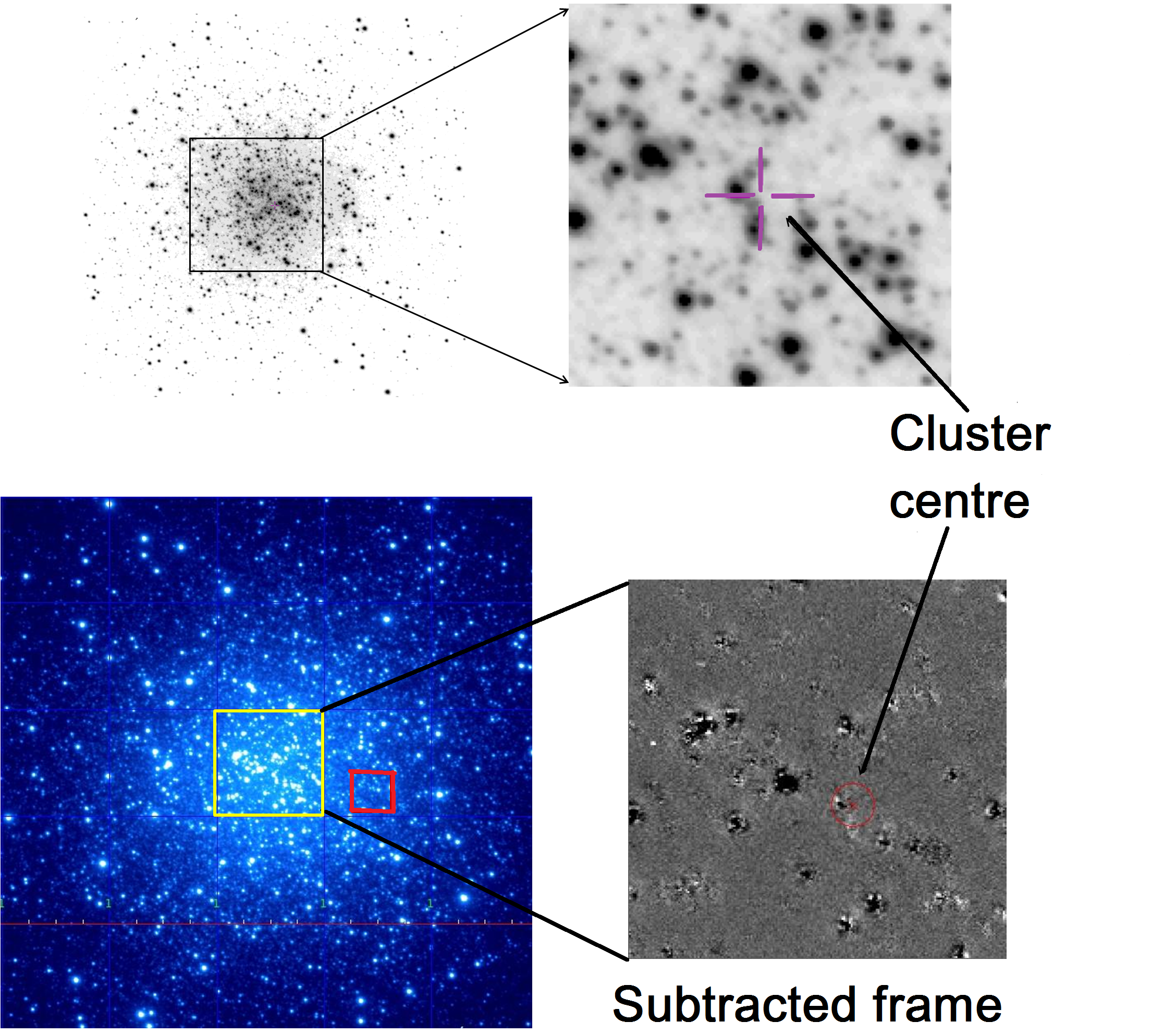}
\caption{M13 RF (left) and subtracted frame (right) of cluster's central part.
\label{fig_1}}
\end{minipage}
\hfill
\begin{minipage}{7cm}
\centering
\includegraphics[width=7cm]{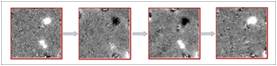}
\caption{A time sequence of zoomed-in part of M13 (corresponding to the small square in RF in Fig.~1) in which we detect 3 known variables with variability of $\sim 0.4$mag: V1, V5, V9 (north to south). 
 \label{fig_2}}
\end{minipage}
\end{figure}
 
\subsection{Searching for free-floating planets (FFPs) with microlensing}

It is now believed that stars with planets are a rule rather than exception, with estimates of the actual number of planets exceeding the number of stars in our Galaxy alone by orders of magnitude (Cassan et al. 2012). However, most of the planets $\sim 4000$ detected till now are in the field of the Galaxy. By a simple estimate, Milky Way's $\sim 200$ globular clusters having $\sim$ million stars each, have a total $\sim 200$ million stars. If there are 10 planets for each star, there shall be $\sim 2$ billion planets in galactic GCs. Actually, there is only 1! -- in a globular cluster M4, discovered accidentally by a pulsar timing. Interesting question arises: where are the GCs planets? It was suggested that in dense GC environment most of the planets will get liberated -- turn `rogue'. The current mass fraction of FFPs may reach $\sim 10\%$ in the entire cluster, and the central density in planets may be $10^5$ pc$^{-3}$ at current epoch, since a significant population of FFPs is retained in relaxed systems (Soker et al. 2001). Their numbers in top 20 dense clusters may even exceed the number of stars by a factor of $\sim 100$ (Fregeau et al. 2002, Hurley \& Shara 2002). FFPs are undetectable by transit or radial velocity methods, and their direct imaging is also not possible. The only way to detect the population of FFPs in GCs is by gravitational microlensing. We have observed GC M4 (chosen due to its proximity, location and the actual existence of a planet) in 2011 in search for ML signatures of FFPs (Safonova et al. 2016). The number of ML events in a time of observation $t^{\rm obs}$ can be calculated considering a
lens (1 Jupiter-mass FFP) with Einstein radius $\theta_{\rm E}$ moving with angular velocity $v$,
\be
\frac{{\Delta N}_{\rm FFP}}{t^{\rm obs}}=\pi \theta_{\rm E}^2 v  \cdot 2 \theta_{\rm GC} \cdot \fr{N_{*}}{4/3\pi\theta^3_{\rm GC}} \times N_{\rm FFP}\,,
\label{self-lensing}
\end{equation}
where $\pi\theta^2_{\rm E}$ is event cross-section, $\theta_{\rm GC}$ is the radius within which we estimate the event rate (e.g. tidal or half-mass), $N_{*}$ is number of stars and $N_{\rm FFP}$ is number of FFPs inside the radius. Velocity dispersion in GCs was shown to depend weakly on mass due to the lack of equipartition (Trenti \& van der Marel 2013), such that $\sigma \propto m^{-0.08}$. In our case, for a mean stellar mass of 1/3 \msun, we estimate for FFPs $\sigma_{\rm FFP} \approx 1.61 \sigma_*$. We assume that velocity vectors have random but isotropic distribution, sources are not moving, and all events have the same characteristic time scale $t_{\rm E}=R_{\rm E}/\s_{\rm FFP}$. For an optimistic stellar density $N_*$ of $10^3$ pc$^{-3}$, the rate of ML events by FFP in 4 months of observations is
\berr
&& N^{\rm low}_{\rm FFP} =5.8\times 10^{-3}\,,\quad \text{where low means } N_{\rm FFP} =N_*\,,\no\\
&&N^{\rm high}_{\rm FFP} =0.58\,,\quad \text{where high means } N_{\rm FFP} = 100\,N_* \,.
\err
Einstein radius of an FFP is about $100$ times smaller than that of central IMBH, and so we expect the duration of ML events to be $\lesssim$ days: e.g. in M4 it is $\sim 19$ hrs for a Jupiter-mass planet.

\section{Summary} 

We propose to exploit the opportunity provided by the large field of view and depth, to study high-redshift sources 
that are magnified by foreground clusters, particularly in the vicinity of the cluster critical curves where we expect 
enhancements of several tens in the brightness of sources that get magnified. Utilizing strong magnification (10--30-fold) of clusters probes much fainter Universe than other methods. For instance, a magnification of roughly 10, combined with angular resolution of $0.^{\as}4$, yields effective resolution of $0.^{\as}04$ in tangential direction. In GCs, the  abundance of  low-mass objects is of particular interest, because it may be representative of the very early stages of star formation in the universe, and therefore indicative of the amount of dark baryonic matter in clusters. 

%
%

%
%
%

\footnotesize
\beginrefer
 
\refer Cassan, A., Kubas, D., Beaulieu, J.-P., et al. 2012, Nature, 481, 167 
 
\refer Ellis R., Santos, M. R., Kneib, J., \& Kuijken, K., 2001, ApJ, 560, L119 

\refer  Fregeau J. M., Joshi K. J., Portegies Zwart S. F. \& Rasio F. A., 2002, ApJ, 570, 171

\refer Hurley, J.~R. \& Shara, M.~M. 2002, ApJ, 565, 1251

\refer Joseph, P., Rozario A., Safonova, M. \& Shastri, P., \#905, Abstract Book 33rd ASI Meeting 2015, Pune, India

\refer Pell\'{o} R.,  Schaerer, D., Richard, J., Le Borgne, J.-F. \& Kneib, J.-P., 2005, RMxAA Conf. Ser, 24, 159 

\refer Pych W., 2012, DIAPL2 at http://users.camk.edu.pl/pych/DIAPL/

\refer Safonova, M. \& Stalin, C.~S., 2010, NewAst, 15, 450

\refer  Safonova, M., Mkrtichian, D., Hasan, P., et al., 2016, AJ, 151, 27

\refer Soker, N., Rappaport \& Fregeau, J. M. 2001, ApJL, 87, 10

\refer Trenti, M. \& van der Marel, R. 2013, MNRAS, 435, 3272 

\refer Wozniak, P.~R. 2000, AcA, 50, 421

\endrefer           

\end{document}